\documentclass[letterpaper,conference]{IEEEtran}
\ifCLASSOPTIONcompsoc
  \usepackage[nocompress]{cite}
\else
  \usepackage{cite}
\fi
\usepackage{comment}
\usepackage{tabularx}
\usepackage{amsfonts}
\usepackage{amsmath}
\usepackage{float}
\usepackage{balance}
\usepackage[T1]{fontenc}
\usepackage{graphics, graphicx,wrapfig}
\usepackage{adjustbox}
\usepackage[caption=false]{subfig}
\usepackage{url}
\usepackage[bookmarks=false, draft]{hyperref}

\usepackage{listings}
\usepackage[T1]{fontenc}
\usepackage[utf8]{inputenc}
\usepackage{multirow}
\usepackage{eqparbox}
\usepackage[table,xcdraw]{xcolor}
\usepackage{listings}
\usepackage{xcolor}
\usepackage{booktabs}
\usepackage{enumerate}

\definecolor{background}{gray}{0.95}
\definecolor{comments}{RGB}{250,167,47}
\definecolor{keywords}{RGB}{1,170,225}
\definecolor{strings}{RGB}{130,204,95}
\lstdefinestyle{codestyle}{
    backgroundcolor=\color{background},   
    commentstyle=\color{comments},
    keywordstyle=\color{keywords},
    numberstyle=\tiny\color{gray},
    stringstyle=\color{strings},
    basicstyle=\ttfamily\small,
    breakatwhitespace=false,         
    breaklines=true,                 
    captionpos=b,                    
    keepspaces=true,                 
    numbers=left,                    
    numbersep=5pt,                  
    showspaces=false,                
    showstringspaces=false,
    showtabs=false,                  
    tabsize=2,
    aboveskip=13px,
    abovecaptionskip=15px
}
\definecolor{ubuntu}{RGB}{85, 17, 64}
\lstdefinestyle{terminal}{
    backgroundcolor=\color{background},    
    commentstyle=\color{comments},
    keywordstyle=\color{white},
    numberstyle=\tiny\color{gray},
    stringstyle=\color{strings},
    basicstyle=\ttfamily\small\color{white},
    breakatwhitespace=false,         
    breaklines=true,                 
    captionpos=b,                    
    keepspaces=true,                 
    numbers=none,                
    showspaces=false,                
    showstringspaces=false,
    showtabs=false,                  
    tabsize=2,
    aboveskip=2px
}

\begin{document}
\title{On-Chain IoT Data Modification in Blockchains}

\author{\IEEEauthorblockN{Sina Rafati Niya, Julius Willems, Burkhard Stiller}
\IEEEauthorblockA{Communication Systems Group CSG, Department of Informatics IfI, University of Zürich UZH\\
Binzmühlestrasse 14, CH---8050 Zürich, Switzerland\\
Emails: [rafati$|$stiller@ifi.uzh.ch], Julius.Willems@uzh.ch}}

\newcommand\eg{\textit{e.g., }}
\newcommand\ie{\textit{i.e., }}
\newcommand\etc{\textit{etc.}}
\newcommand\cf{\textit{cf. }}

\maketitle

\begin{abstract} 
In recent years, the interest growth in the Blockchains (BC) and Internet-of-Things (IoT) integration --- termed as BIoT--- for more trust via decentralization has led to great potentials in various use cases such as health care, supply chain tracking, and smart cities. A key element of BIoT ecosystems is the data transactions (TX) that include the data collected by IoT devices.  BIoT applications face many challenges to comply with the European General Data Protection Regulation (GDPR) \ie enabling users to hold on to their rights for deleting or modifying their data stored on publicly accessible and immutable BCs. 
In this regard, this paper identifies the requirements of BCs for being GDPR compliant in BIoT use cases.

Accordingly, an on-chain solution is proposed that allows fine-grained modification (update and erasure) operations on TXs' data fields within a BC. The proposed solution is based on a cryptographic primitive called Chameleon Hashing.
The novelty of this approach is manifold. BC users have the authority to update their data, which are addressed at the TX level with no side-effects on the block or chain. By performing and storing the data updates, all on-chain, traceability and verifiability of the BC are preserved. 
Moreover, the compatibility with TX aggregation mechanisms that allow the compression of the BC size is maintained.

\end{abstract}
\IEEEpeerreviewmaketitle

\section{Introduction}
\label{sec:intro}

With the increase of user data misuse by third parties, international society has realized the potential threats imposed by the lack of user privacy regulations. In 2018, the European Union (EU) responded to this public concern by enforcing the European General Data Protection Regulation (GDPR), which regulates the relationship between a user (data subject) and an organization (service provider) that stores any form of the users' personal data. GDPR intends to grant substantial rights to data subjects. Especially, by declaring article 16 \cite{GDPR16} on ``right of rectification" and article 17 \cite{GDPR17} on ``right to be forgotten", GDPR empowers individuals to request the ``immediate deletion" or ``immediate rectification" of their personal data. In case of a violation, substantial fines would apply to the responsible service provider. GDPR applies to any service provider storing or processing the personal data of individuals residing in the EU. This translates to a universal application of the new law. For example, a US-based video-streaming company serving EU customers needs to comply with GDPR.

Blockchains (BC) are distributed data storage systems, There are no commonly agreed definitions but a blockchain is often considered a special kind of a distributed ledger (DL) with records bundled in (chained) blocks. BCs store the data within backward linked blocks, and verify the data integrity of blocks using hash functions. 

A hash function \(h: \{{0,1}\}^* \xrightarrow{} \{{0,1}\}^k\) maps an input string \(a \in \{{0,1}\}^*\) of arbitrary length to an output string \(b \in \{{0,1}\}^k\) of fixed length \(k\) such that
\(\forall a \in \{{0,1}\}^* \exists b \in \{{0,1}\}^k \wedge k\in \mathbb{N}\)

One key characteristic of hash functions \(h: \mathbb{A} \xrightarrow{} \mathbb{B}\) is their resilience to collisions in the co-domain. This means it is hard to find two \textit{distinct} values with the same hash output.

\begin{equation}
    a,b \in \{{0,1}\}^* \wedge{} a\neq{}b \wedge{} h(a)=h(b)
\end{equation}

BCs, like many other applications depend on the collision-resistance of cryptographic hash functions. Using cryptographic hash functions, the chain of blocks --\ie blockchain-- is persisted in a tamper-proof fashion. Lets assume a block as a tuple \(B=(p, c, n)\) defined as follows.
\begin{itemize}
    \item \(p \in \{{0,1}\}^k\) is a pointer of fixed length \(k \in \mathbb{N}\) to the previous block
    \item \(c \in \{{0,1}\}^*\) is a stream of 0s and 1s representing the content of a block
    \item \(n \in \mathbb{N}\) is a special number associated with the block's content.
\end{itemize}

An important property of \(n\) is the following: When concatenated with the block's content \(c \mathbin\Vert n\) and then hashed, the resulting hash has a prefix of \(0s\) of fixed length \(i\). Due to the collision-resistance of cryptographic hash functions, a suitable \(n\) for a given \(c\) can only be found by brute force and is computationally expensive. Whenever such an \(n\) is found, the Proof-of-Work puzzle is solved and the block is valid:
\begin{equation}
    valid(B) \iff g(c \mathbin\Vert n) = 0_{1}0_{2}0_{3}...0_{i}\mathbin\Vert \{{0,1}\}^{k-i}
\end{equation}
Where \(f, g:\{{0,1}\}^* \xrightarrow{} \{{0,1}\}^k\) are hash functions and \(\mathbin\Vert\) denotes the concatenation operation. 

Let \(C\) be a chain of blocks with \(B_n=(p_n, c_n, n_n)\) being the last block. A new block \(B_{n+1}=(p_{n+1}, c_{n+1}, n_{n+1})\) can be attached to \(B_n\) such that \(p_{n+1}=f(p_n \mathbin\Vert g(c_n \mathbin\Vert n_n))\). This yields the new chain \(C'=B_{1}\mathbin\Vert B_{2}\mathbin\Vert...\mathbin\Vert B_{n}\mathbin\Vert B_{n+1}\). It becomes clear that altering the content of block \(B_i\) from \(c_i\) to \(c_{i'}\) yields \(p_{i+1} \neq{} f(p_i \mathbin\Vert g(c_{i'} \mathbin\Vert n_i))\). This break propagates onwards and renders all subsequent blocks invalid. It is important to observe here that the collision-resistance property (\cf Equation (\textit{1)}) of \(g(\mathbf{c_{i'}} \mathbin\Vert n_i) \neq{} g(\mathbf{c_{i}} \mathbin\Vert n_i)\) is responsible for the diverging output and thus the break of the chain.

BCs have gained popularity ranging from Fintech and De-Fi (Decentralized Finance) \cite{trademap} to Supply Chain Tracking \cite{DairySCT}, Health care, and smart city monitoring via Internet-of-Things (IoT) devices. Integration of BCs and IoT (BIoT) has shown  potential in such use cases mostly due to the BCs' immutability and tamper-proofness. However, BCs face many challenges, such as scalability, security, and privacy \cite{Noms20, lcn19}.

BCs data storage is distributed among its users, especially the miners who validate the transactions (TXs) and mine the blocks. Any data stored within a TX, once added into a BC, is accessible by all BC nodes. These nodes can -- and even in some cases, they have to -- store the full chain locally. Hence, a great challenge BIoT applications face is the coexistence of BCs with GDPR when someone's personal data is inserted into a BC. An example scenario can be the storage of health status data collected in a clinical study in a BC. Due to the public access to that data stored on the BC, user privacy is endangered. To address such user-centric privacy issues, BIoT platforms are interested in complying with the GDPR. However, the coexistence of BCs and GDPR is highly challenging due to the immutability of BCs. 

The question raises is how can the ``Rights to be forgotten and rectification" and the BCs immutability coexist? To address this question, this work presents a novel approach that adapts a Proof-of-Stake (PoS)-based BC namely Bazo, to an updatable BC. Bazo is a public-permissioned BC, designed for IoT-oriented use cases \cite{Bazo-ACM, bazo-demo}. To adhere to the GDPR, this work's approach enables transparent and traceable data modification at the TX level. For that purpose, a particular hash function, \ie Chameleon Hash Function (CHF) is employed.

The remainder of this paper is organized as follows. Section \ref{sec:RelatedWork} presents the background and related work. The design and implementation of the proposed approach is described in Section \ref{sec:deisgn} and the evaluation discussion is presented in Section \ref{sec:Eval}. Finally, Section \ref{sec:summary} covers a brief summary.

\section{Related Work}
\label{sec:RelatedWork}

Approaches to complying BCs with GDPR are very recent and fundamentally different. This work conducted a study on the state of the art approaches that yield two identified categories \ie structural and cryptographic elaborated as follows. 

\subsection{Structural Approaches}

\subsubsection{Off-chain storage} \label{offchain}
One approach to harmonize BCs with GDPR, is the application of off-chain storage \cite{eberhardt, garcia}. \cite{eberhardt} proposes a solution where the storage of data is outsourced to an external, off-chain data storage. To this end, the way TXs are stored into the BC differs from traditional BCs. The TX's data-payload is replaced with the hash of the original data  \(hash = h(data_{orig})\). The \(data_{orig}\) is then stored in a trusted, off-chain data storage. This data storage is a key-value store, and takes \(hash\) as key and \(data_{orig}\) as value.

Whenever a participant queries the data of a TX, it uses \(hash\) which serves as key for the external data storage. Upon reception of the external \(data_{ext}\), the client can assume the data is untampered whenever \(h(data_{orig}) = h(data_{ext})\) holds.
When a data subject requests its right to be forgotten, the individual or organization in charge of the central data storage performs a modification on the data. Due to the fact that data is being stored off-chain, a deletion doesn't affect the content of a TX and thus doesn't affect the BC's consistency \cite{eberhardt}. 

Considering the current state of legal uncertainty, this defensive approach aims to avoid tension with GDPR by outsourcing the issue to a proven technical solution. Therefore, it is also one of the most popular solutions applied by the industry so far \cite{eberhardt}. Enabled by the outsourced data, the BC size is reduced which is clearly a positive side-effect. Additionally, deletions and modifications can be performed on a TX-level which allows fine-grained control.

The biggest constraint in this approach is the centralized data storage. It compromises one of the defining characteristics of BCs - distribution of data. While minimizing data storage on the BC is in line with GDPR, data hash, even after modification, renders a risk-factor for non-compliance. The hash constitutes, when combined with other information, pseudonymous data, which doesn't meet the standard of anonymous data as required by the GDPR \cite{bc_solutions}.

\subsubsection{Storing TX Hashes} \label{omit_content}
Omitting the TX content is another approach where no off-chain storage is required. For instance, \cite{mof_bc} aims to reduce the overall BC size by employing a reward system that incentivizes BC users to compress or delete their data. \cite{mof_bc} introduces multiple agent roles assigned to individual miners to handle the different tasks associated with compression, deletion, and rewards distribution. Thus, a centralized entity is responsible for auditing and verifying the agents mentioned above. 

To reduce storage requirements and enhance privacy, \cite{mof_bc} developed a new type of TX \(dTX\) that incorporates deletion of a TX \(TX_i\). It is designed in a way that only the creator of the TX \(TX_i\) can request the deletion of the same TX. For this purpose, \(dTX\) features an input field holding the TX-id \(TX_i.id\). If a TX creator wants to delete it, he generates \(dTX\) and signs it with the same private key as \(TX_i\) was signed. Using a signing scheme, \cite{mof_bc} verify that the originator of \(TX_i\) and \(dTX\) are identical. 

\(dTX\) is then broadcasted to the network of miners. Upon receipt, a miner first has to query \(TX_i\) by the given \(TX_i.id\). Querying \(TX_i\) is performed in \(O(N)\) where \(N\) is the number of TXs in the BC. Since \(N\) typically grows over time, this linear scan yields a potential slowdown of the BC. To avoid performance degeneration, miners locate TXs asynchronously in a so called Cleaning Period (CP). The length and frequency of the CP is based on the application. The main idea behind a CP is, however, that the miners perform the deletions in batches (one after the other) to increase efficiency. The authors also propose that the batch-removal can be performed by one miner only who then broadcasts the updated chain to the network. Nonetheless, other miners can still perform the batch deletion on their own and compare their results to ensure data integrity. 

In order not to break hash consistency among blocks when deleting TXs, the block and TX structure has to be adjusted. In conventional BCs the hash of the block includes the Merkle root of those TXs \[MerkleRoot = MerkleTree(TX_1 \mathbin\Vert TX_2 \mathbin\Vert ... \mathbin\Vert TX_n)\] where \(TX_i\) is a single TX. To mitigate breaking hashes, \cite{mof_bc} proposes a design where the Merkle root is generated over the stored hashes of the TXs  \[MerkleRoot = MerkleTree(TX_1.h \mathbin\Vert TX_2.h \mathbin\Vert ... \mathbin\Vert TX_n.h)\] where \(TX_i.h\) is the hash of \(TX_i\). Note that \(h\) is a \textit{field} on \(TX\). This means that \(h\) is computed once when the TX is created such that \(TX_i.h = h(TX_i)\). Consequently, during the calculation of the Merkle root, the TX-hash is looked up rather than calculated during runtime. A deletion is then performed by clearing all fields on the TX except \(TX.h\), ensuring consistent Merkle roots and thus block-hash consistency.

The absence of a centralized database --- as opposed to the previous solution --- is an improvement in \cite{mof_bc}. Additionally, the modification is possible on a TX-level, which enables fine-grained control. However, this approach can be criticized for centralized agents managing different tasks, thus, limiting its application to authorized nodes in a permissioned BC only. Further, it is unclear whether the delay of a deletion contradicts with the GDPR requirement of an immediate deletion. Lastly, storing the hashes on TXs compromises verifiability as a malicious participant could modify fields of the TXs without them standing out in the Merkle root computation.

\subsubsection{Replacing Blocks} \label{replace_blocks}
Replacing entire blocks goes in a similar direction as omitting the TX content. The protocol developed in \cite{bc_permissionless_redactable} accommodates modifications in BCs via democratic voting as a means for mutating the BC state. Every BC client can propose an edit operation, whether it may be a deletion or modification of data. Therefore, BC miners vote on whether to accept or reject the change.  \cite{bc_permissionless_redactable} stores the Merkle root of the block in two fields. \ie the \textit{old state}, and the \textit{Merkle root}. Which means a block \(B_i\) is connected to its predecessor \(B_{i-1}\) by 2 links:
\begin{enumerate}
    \item \(B_{i}.PrevOldState = B_{i-1}.OldState\)
    \item \(B_{i}.Prev = B_{i-1}.MerkleRoot\)
\end{enumerate}

When \(TX_{j}\) stored in block \(B_{i}\) needs to be edited,  a new block \(B_{i'}\) is created such that:
\begin{enumerate}
    \item \(B_{i'}.PrevOldState = B_{i-1}.OldState\)
    \item \(B_{i'}.Prev = B_{i-1}.MerkleRoot\)
    \item \(B_{i'}.OldState = B_{i+1}.PrevOldState\) 
    \item \(B_{i'}.MerkleRoot \neq B_{i+1}.Prev\)
\end{enumerate}

(1) and (2) remain unchanged due to changes in a block are propagated to subsequent blocks only. (3) is enabled by the added field \textit{old state} and ensures that at least one valid link exists between \(B_{i+1}\) and \(B_i\). (4) is the consequence of the change. Miners can now vote about the change. If a majority is reached, \(B_{i}\) is swapped with \(B_{i'}\).

This approach relies on storing the hash of the original state to preserve hash consistency. The absence of a centralized agent and the application democratic voting can be seen as an improvement over the previous approach. However, this approach focuses on replacing entire blocks and thus does not allow fine-grained modifications on TX-level. Similar to \cite{mof_bc}, the hash of the data is stored rather than computed, resulting in compromises in verifiability.

\subsection{Chameleon Hash Function (CHF)} \label{cham}
CHF compliments the cryptographic primitives of BCs. A CHF is a special kind of hash function that makes use of the ``collision" concept \cite{ch_sig}. A CHF is associated with a public key (pk) and private key. The private key is also referred to as the trapdoor key (tk). Without \(tk\) the hash function is collision-resistant. Using \(tk\), it is computationally possible to find an input \(a'\) for an already existing input \(a\) where \(a \neq a'\)  and \(h(a) = h(a')\) \ie generating a hash collision.

A CHF consists of a set of four algorithms \(CHF=\{H_{trp}, H, H_{ver}, H_{col}\}\) defined as follows \cite{bc_friends,ch_collision_resistant}.

\begin{enumerate}[i]
\item \((pk, tk) \xleftarrow{} H_{trp}(1^k)\) is an algorithm given a security parameter \(k \in \mathbb{N}\) to compute the public hash key \(pk\) and a secret trapdoor \(tk\).
\item \((h, \xi) \xleftarrow{} H(pk, m)\) is an algorithm given the public hash key \(pk\) and a message from the set of possible messages \(m \in \mathbb{M}\) that returns a hash value \(h\) and a check string \(\xi\) derived from \(m\).
\item \([0,1] \xleftarrow{} H_{ver}(pk, m, (h, \xi))\) is an algorithm given the public hash key \(pk\), a message \(m\) and a tuple of a hash \(h\) and a check string \(\xi\) that returns \(1\) iff the tuple \((h, \xi)\) is valid for the message \(m\) and the pulibk key \(pk\), otherwise it returns 0. 
\item \(\xi' \xleftarrow{} H_{col}(tk, (h, m, \xi), m')\) is an algorithm that takes the trapdoor \(tk\), a valid tuple \((h, m, \xi)\) and a message \(m' \in \mathbb{M}\) with \(m 
\neq{m'}\) and returns a new check string \(\xi'\) such that \( H_{ver}(pk, m', (h, \xi')) = H_{ver}(pk, m, (h, \xi)) = 1\) holds.
\end{enumerate}

\cite{bc_friends} uses CHF to address the issue of breaking hashes outlined above. To this end, let \(g=\{H_{trp}, H, H_{ver}, H_{col}\}\) be a CHF. Further, a block is extended to \(B=(p, c, n, (h, \xi))\). The additional tuple \((h, \xi) \xleftarrow{} H(pk, (\underbrace{p, c}_\text m))\) represents a hash/check-string combination generated by \(ii\). The block is considered valid if the following condition exists. 
\begin{equation}
    valid(B) \iff H_{ver}(pk, (\underbrace{c, p}_\text m), (h, \xi)) = 1
\end{equation}
 Let \(C\) be a chain of blocks with \(B_n=(p_n, c_n, n_n, (h_n, \xi_n))\) as the last  block. A new block \(B_{n+1}=(p_{n+1}, c_{n+1}, n_{n+1}, (h_{n+1}, \xi_{n+1}))\) to \(B_n\) can be attached now such that \(p_{n+1}=f(h_n \mathbin\Vert n_n)\) with \(h_n\) as the chameleon hash of the block's content. This yields the new chain \(C'=B_{1}\mathbin\Vert B_{2}\mathbin\Vert...\mathbin\Vert B_{n}\mathbin\Vert B_{n+1}\).

The content of a block \(B_i=(p_i, c_i, n_i, (h_i, \xi_i))\) can be changed from \(\mathbf{c_i}\) to \(\mathbf{c_{i'}}\) by the party in possession of \(tk\). To this end, first a new check string using \((iv)\) has to be generated:
    \[
    \xi_{i'} \xleftarrow{} H_{col}(tk, (h_i, (\underbrace{c_i, p_i}_\text m), \xi_i), (\underbrace{\mathbf{c_{i'}}, p_i}_\text {m'}))
    \]

The new content \(\mathbf{c{i}'}\) and check string \(\mathbf{\xi_{i'}}\) are set on the block such that \(B_i=(p_i, \mathbf{c_{i'}}, n_i, (h_i, \mathbf{\xi_{i'}}))\) with the effect that:
\begin{equation}
    H_{ver}(pk, (\underbrace{\mathbf{c_i}, p_i}_\text m), (h_i, \mathbf{\xi_i)}) = H_{ver}(pk, (\underbrace{\mathbf{c_{i'}}, p_i}_\text {m'}), (h_i, \mathbf{\xi_{i'}})) = 1
\end{equation}

One can now observe that \(h_i\) and \(n_i\) remain the same before and after the redaction. Thus, the link in the subsequent block \(p_{i+1}=f(h_i \mathbin\Vert n_i)\) remains valid \cite{bc_friends}.

\cite{bc_friends} focuses on the cryptographic aspect of the BC, rendering a centralized authority that stores data or manages roles. Also, the fact that the content can be replaced completely (\(m\) and \(m'\)) allows a strict realization of the GDPR's requirement of erasure. However, similar to the approach in section \ref{replace_blocks}, it suffers from the lack of fine-grained control as the modification is done on a block-level. Additionally, the management of the trapdoor key as to which individuals are allowed to perform modifications imposes new challenges \cite{bc_solutions}. Succeeding works, like the one from \cite{att_cham}, have developed novel solutions to overcome the challenge of trapdoor key management and access control.

\begin{table*}[htbp]
\caption{\label{table:related-work} Comparative Overview on Potential GDPR-compliant Approaches}
\resizebox{\textwidth}{!}{%
\begin{tabular}{|l|l|l|l|l|}
\hline
\textbf{Approach}                                                            & \textbf{Description}                                                                                                                        & \textbf{\begin{tabular}[c]{@{}l@{}}Level of\\ Modification\end{tabular}} & \textbf{Advantages}   & \textbf{Disadvantages}                                                                                            \\ \hline
Off-chain Storage~\cite{eberhardt}                                                            & \begin{tabular}[c]{@{}l@{}}Data is stored \\off-chain in a \\centralized database\end{tabular}                                                 & Transaction                                                              & \begin{tabular}[c]{@{}l@{}}Minimal tension \\ with GDPR\end{tabular}                                                 & \begin{tabular}[c]{@{}l@{}}1- Compromises \\ distribution of data.\\2- Dependency on \\external system\end{tabular}                              \\ \hline
\begin{tabular}[c]{@{}l@{}}Storing transaction hashes~\cite{mof_bc}\end{tabular}       & \begin{tabular}[c]{@{}l@{}}Hash of transaction is\\ looked up rather\\ than computed\end{tabular}                                               & Transaction                                                               & \begin{tabular}[c]{@{}l@{}}Simple approach \\low complexity\end{tabular} & \begin{tabular}[c]{@{}l@{}}1- Compromises traceability \\ 2- Compromises auditability\end{tabular}                     \\ \hline
Replacing blocks~\cite{bc_permissionless_redactable}                                                             & \begin{tabular}[c]{@{}l@{}}Miners vote \\on replacing blocks\end{tabular}                                & Block                                                                     & \begin{tabular}[c]{@{}l@{}}Respects a voting policy \end{tabular}                                         & \begin{tabular}[c]{@{}l@{}}Only possible to replace\\ entire blocks\end{tabular}                         \\ \hline
Chameleon hashing~\cite{bc_friends}                                                            & \begin{tabular}[c]{@{}l@{}}Hash collisions \\can be used to\\ replace content\end{tabular} & Block                                                                     & \begin{tabular}[c]{@{}l@{}}Maintains traceability\\ and auditability\end{tabular}                                    & \begin{tabular}[c]{@{}l@{}}1- Complexity of cryptography\\2- Storage of trapdoor key\end{tabular} \\ \hline
\begin{tabular}[c]{@{}l@{}}Attribute-based \\ chameleon hashing~\cite{att_cham}\end{tabular} & \begin{tabular}[c]{@{}l@{}}Adds fine-grained \\ access control to \\ chameleon hashing\end{tabular}                                          & Transaction                                                               & \begin{tabular}[c]{@{}l@{}}Enables a fine-grained \\ control as to who \\ is able to perform an update\end{tabular}  & \begin{tabular}[c]{@{}l@{}}Access scheme adds an\\ additional layer of\\complexity\end{tabular}      \\ \hline
\end{tabular}}
\end{table*}

\subsubsection{Policy-based Chameleon Hashing} \label{attr_cham}
\cite{att_cham} improves the work in \cite{bc_friends} by developing a cryptographic permission scheme enabling more fine-grained control over BC modifications. \cite{att_cham} proposes an attribute-based access control to implement policies on whether a user is authorized to perform a modification. To this end, boolean formulas are evaluated over a set of user attributes. If the user is in possession of \(tk\) and satisfies the boolean formula, he or she is authorized to perform the modification.
Consequently, the notion of policy-based Chameleon Hashing (PCH) is introduced which extends chameleon hashing by additionally taking an access policy as an input parameter. PCH works in a way that those hash collisions can only be generated by users satisfying the access policy and in possession of \(tk\). The access policy can be defined by the user generating the Chameleon Hash of a resource.

For example, two users are associated with attributes \(user1=\{C\}\), \(user2=\{B, D\}\) and a resource has the access policy \((A \wedge{} B) \vee C\). While \(user1\) would be authorized, \(user2\) does not satisfy the required access policy. Basically, any combination of attributes and logical operators can be used to generate an access policy.

 Moreover, \cite{att_cham} offers the key management for users. In the approach of \cite{bc_friends} the key-pair \((pk, tk)\) is created and associated on a per user basis. As a consequence, the user can find hash collisions for all hashes generated by his/her \(tk\). Losing \(tk\) or if the \(tk\) was stolen, a high security risk could be raised. To mitigate this risk, \cite{att_cham} generates a new pair of \((pk, tk)\) for every generated hash. Hence, an advantage over the previous solution is made possible by modifying the BC content on a TX-level -- allowing more fine-grained control. However, the improved access-control adds an additional layer of cryptographic complexity to the already complex chameleon hashing scheme. Another disadvantage is using a different key-pair for each hash also causes a demand for a more sophisticated key storage. 

\subsection{Discussion and Requirements Specification } \label{discussion}

These different approaches are compared in Table~\ref{table:related-work}. Based on these studies, the following seven factors are derived as high-priority requirements for a GDPR-Complaint BC development to be considered in BIoT-based systems.
\begin{enumerate}
    \item Modifications shall enable the updates or deletion of personal/IoT data stored in the BC.
    \item A user shall only be able to request the modification of his/her own personal/IoT data.
    \item Modifications shall be performed immediately and at worst case, in less than 24 hours.
    \item The BIoT-based system cannot depend on external services/systems (\eg external Database) for modification.
    \item Modifications shall be possible in a fine-grained way.
    \item Modifications on the BC shall be transparent, recorded, and traceable.
    \item Modifications shall not affect existing BC optimizations such as TX aggregation 
\end{enumerate}

\section{Design and Implementation}
\label{sec:deisgn}
\begin{figure*}[ht]
    \centering
    \includegraphics[width=0.8\textwidth, height=0.27\textheight]{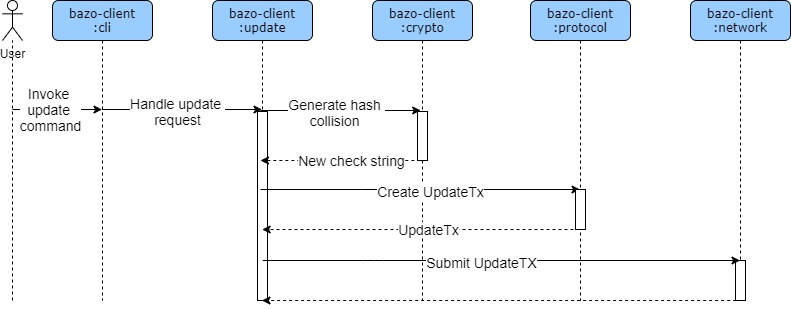}
    \caption{Client Interactions with Bazo for Updating a Transaction}
    \label{fig:client_sequence}
\end{figure*}

The approach in this work adapts the Bazo BC to a GDPR-complaint version of it by considering the seven requirements discussed in Section \ref{discussion}. While the rest of this Section elaborates on the design and implementation of the approach taken in this work, Figures \ref{fig:client_sequence} and \ref{fig:miner_sequence} represent the steps taken at BC client and miner sides for deletion and modification processes -- \ie update processes. 

\subsection{Data Ingestion and Storage} \label{data_ingestion}
As defined by requirement \textit{(1)}, it shall be possible to remove or modify personal data stored in the BC. With the absence of legal precedence as to what exactly qualifies as personal data \cite{gdpr_bc_eu_study}, the scope and implications of a modification are not clear. As a consequence, an assumption is made in this work in order to keep the design flexible to compensate for this uncertainty. Hence, an abstraction layer is introduced in this work to hide the ambiguity of personal vs. non-personal data. 

The same as in most other BCs, in Bazo data is ingested by several types of TXs each with different functionality (\eg account TXs and funds TXs). In this work, a TX's data structure is chosen to implement the abstraction mentioned above. Hence, TXs initiated by users or IoT devices, are extended with a new generic field called ``Data" to hold (potential) personal data, subject to GDPR. Other fields of a TX (\eg \(Amount\) or \(Fee\)) do not render personal data. In this work, only content of the \(Data\) field is considered as personal data. Users are able to pass along data with every TX they emit. Their data will be stored in the designated \(Data\) field in the TX to be broadcast among the network of miners and eventually added to a block. 

Thus, the requirement \textit{(1)} \ie modification right, is mapped to the \(Data\) field on TX data structure. Which means, to implement a modification request, the designed approach here updates only the \(Data\) field.

\subsection{Replacing the Traditional Hash Function}
If the \(Data\) field of a TX is updated due to a modification request, that TX's hash changes, which propagates further causing a different Merkle root of the block the TX was included in. As a consequence, the change of a single TX hash would eventually break the hash consistency of the entire chain in a BC. While this is the intended behavior for other fields on the TX (\eg \(Amount\) field of a funds TX), modifications of the \(Data\) field can have a different behavior.

Requirement \textit{(2)} implies that only the user who created the TX and passed personal data into a BC should be able to alter the same data without changing the hash of the TX. Evenly important, any other user shall not be able to change the \(Data\) field of that TX without changing its hash. To address this requirement the SHA-3 hash function used previously in Bazo is replaced with a CHF.

With the approach in this work, whenever a new client account is issued, a set of CHF parameters is generated alongside. Thus, Bazo clients are modified to use the CHF parameters to hash the TXs they emit. As generating a hash for an input value requires a public hash key (\cf Section \ref{cham}), and returns the digest together with a check string. The digest can then be recomputed with the input value, public hash key, and check string. As the CHF signature and return value differ from the hash function used in Bazo, a drop-in replacement is not possible. 

For BC participants, in order to be able to hash a certain TX, the public hash key and check string need to be publicly available. To achieve this characteristic, data structure adjustments were made on Bazo. Accordingly, all data embedded TXs are extended with a field to store their generated check string. Moreover, Bazo client (user/IoT device) accounts are extended to store the CHF parameters generated in the account creation step. This ensures that a Bazo miner is now able to compute the hash of a TX as it uses the check string stored on the TX and the client's CHF parameter stored in the account. With these updates, each Bazo client has it's own set of CHF parameters and uses them to hash his/her/its TXs.

\subsection{CHF Specification}
The CHF parameters developed in this work define a set of 5 numbers as follows \cite{CHF-code}.

\begin{verbatim}
type ChameleonHashParameters struct {
    G  []byte // Prime
    P  []byte // Prime
    Q  []byte // Prime
    HK []byte // Public Hash Key
    TK []byte // Secret Trapdoor Key
}
\end{verbatim}

where \(G, P, Q\) are special prime numbers created by the parameter generation algorithm. \(HK\) is the public hash key required to compute the CHF of an input value. \(TK\) is the secret trapdoor key enabling the generation of hash collisions. Just like the secret key of an RSA key-pair, it should be stored safely and never shared with others.

The designed approach in this work is implemented as follows. The CHF parameters are created during the account creation process and written to a file on the client's machine. As the CHF parameters also contain the public hash key, they need to be accessible for other entities as well. In particular, miners need to recompute TX hashes and thus need access to the client's public hash key. As the miners in Bazo have access to all BC accounts, the CHF parameters are attached to the client's account and included in the \verb|AccountTx|. However, before they are attached, the secret \textit{tk} is sanitized. When the \verb|AccountTx| is sent to the network and the account is created, all miners have access to the client's public hash key.

\begin{verbatim}
parameters.TK = []byte{} 
\end{verbatim}

\begin{verbatim}
type AccountTx struct {
    ...
    Parameters  *crypto.CHFParameters 
    CheckString *crypto.CHFCheckString 
    Data        []byte
}
\end{verbatim}

When a client issues a new TX it creates a TX object, fills the fields of the TX with the given values, signs the TX, and submits it to the network of miners. The signature consists of the TX hash, signed by the client's private key. Before shifting to chameleon hashing, the client application used a conventional SHA3 hash function to compute the TX hash. The message (input to the hash function) incorporates all fields of the TX. To that end, each TX implements a hash method.

An example of \verb|FundsTx| \verb|Hash()| method:

\begin{verbatim}
func (tx *FundsTx) Hash() [32]byte {
 input := struct {
     Header byte
     Amount uint64
     Fee    uint64
     TxCnt  uint32
     From   [32]byte
     To     [32]byte
     Data   []byte
   }{ tx.Header, tx.Amount, tx.Fee, 
     tx.TxCnt, tx.From, tx.To, tx.Data }
return sha3.Sum256(
[]byte(fmt.Sprintf("%v", input)))}
\end{verbatim}

As mentioned in Section \ref{cham}, in contrast to the traditional hash functions \eg SHA2 and SHA3 where hash generation and verification are done by re-/computing the hash, CHF does the hash generation and verification using two separate procedures. For that, the client needs to generate an initial check string. 
\begin{verbatim}
type ChameleonHashCheckString struct {
    R []byte
    S []byte }
\end{verbatim}

The check string is generated using the CHF parameters of the client. \(R, S\) are random numbers with an upper bound defined by the CHF's \(Q\) parameter. 

\begin{verbatim}
parameters := crypto.NewCHFParameters()
checkString := crypto.NewCHFCheckString(
    parameters)
\end{verbatim}

Afterwards, parameters, check string and message (hash input) are used to compute the 32-byte CHF.

\begin{verbatim}
chameleonHash := crypto.CHF(
parameters, checkString, message)
\end{verbatim}

The message in our case is the SHA-hash of the TX we saw above. To keep check strings organized, they are stored directly on the TX they belong to. To that end, each TX is extended with a \verb|CheckString| field.

\subsection{Performing a Modification} \label{performing_a_modification}
With the changes described in the previous section, prerequisites are met for BC clients to perform modifications. Firstly, a Bazo client chooses one of its TXs to be modified. Subsequently, the TX's \(Data\) field can be modified (either cleared or overwritten). Lastly, the client needs to generate a hash collision for the contents of the original and modified TX. Generating a hash collision requires the secret \textit{tk}, the old hash input, the new hash input, and the old check string. CHF parameters are issued on a per account basis. As the secret \textit{tk} always stays with its owner, only clients can generate hash collisions for their own TXs. Generating a hash collision returns a new check string associated with the modified TX content. When a new TX is hashed including [new data / public hash key / new check string], the same digest is returned as before the modification with [old data / public hash key / old check string]. At this point, the client successfully modified the \(Data\) field of a TX and generated a hash collision. An attempt to generate a hash collision for a TX that does not belong to the client (\eg the check string on the TX was not created by the client's CHF parameters) would fail in the sense that the resulting hashes would not match.

\subsection{UpdateTx} \label{update_tx}
Until now, the scope of modifications described in the previous section was limited to a client's local machine. In order for the modification to come into effect, updated TXs need to be sent to all the distributed BC miners and applied to their local copy of the BC. In that context, the notion of \(UpdateTx\) is introduced in this work as a new TX type in Bazo BC. An \(UpdateTx\) is serving as a carrier for the modification. The motivation for using TXs to request data modification consists of two reasons. Firstly, as identified in section \ref{data_ingestion}, TXs are the only way to interact and alter the state of a BC, thus this concept is used to modify the Bazo BC as well. Secondly, TXs provide auditability and traceability. Requirement (6) (\cf Section \ref{discussion}) states that BC participants shall be able to trace any modifications performed on the BC. Thus, an \(UpdateTx\) contains all the information about the modification on the BC and is mined in a block -- just like any other TX. Evidently, they record the modifications transparently and traceable.

The \(UpdateTx\) specifies the following fields:
\begin{itemize}
    \item \textbf{TxToUpdateHash} The hash of the TX a client wants to modify.
    \item \textbf{TxToUpdateCheckString} The new check string the client computed to generate a hash collision.
    \item \textbf{TxToUpdateData} The modified data to be set on \(TxToUpdate\).
    \item \textbf{Issuer} The account address of the client.
    \item \textbf{CheckString} The check string for this \(UpdateTx\).
    \item \textbf{Data} The data field of this \(UpdateTx\).
    \item \textbf{Signature} The signed hash of this \(UpdateTx\).
\end{itemize}

The client signs the \(UpdateTx\) with his/her private key and sends the TX to the network of miners.

\subsection{Validation of an UpdateTx} \label{validation_of_update_tx}

\begin{figure*}[ht]
    \centering
    \includegraphics[width=0.95\textwidth, height=0.4\textheight]{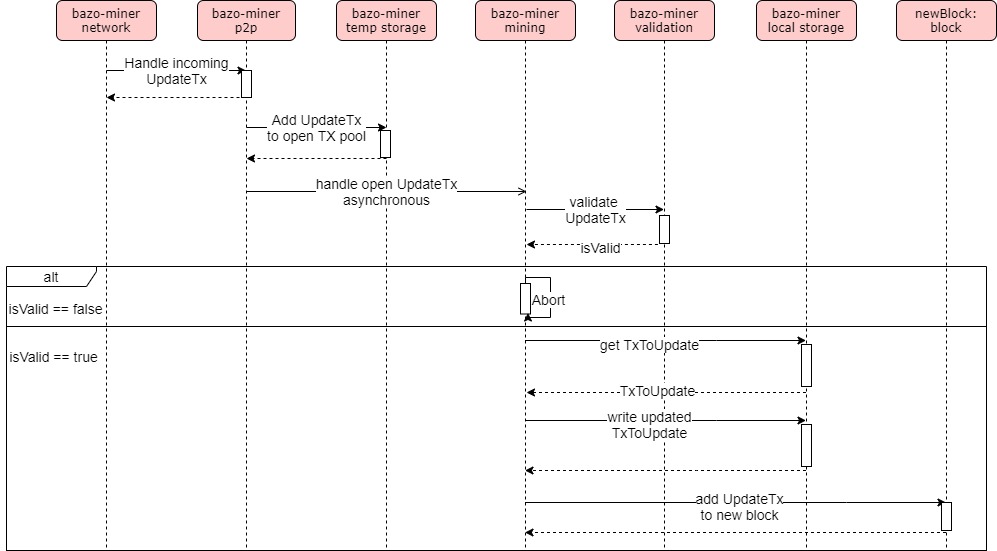}
    \caption{Miner Interactions in Bazo for Processing an update Transactions}
    \label{fig:miner_sequence}
\end{figure*}

When a miner receives a new TX it adds the TX to a set of open TXs (\cf Figure \ref{fig:miner_sequence}). When a new mining round starts, the miner validates, processes and adds the TXs to the a new block. The same is applied to \(UpdateTx\) as well. In this step, the miner performs checks to ensure the client is authorized to request the modification. In the first step, the miner verifies if \(TxToUpdate\) exists. For that matter, the miner queries the \(TxToUpdate\) in its local storage. If the TX doesn't exist, the process is terminated. Next, the miner checks whether the client's account exists in the BC. Upon success, miner verifies that the signature on the \(UpdateTx\) was signed with the private key corresponding to the client's public key. If the signature is valid, the miner compares if the \(TxToUpdate\) was signed with the same private key as the \(UpdateTx\). This check verifies that clients can only update their own TXs. As pointed out in Section \ref{performing_a_modification}, this constraint is already implemented by the distribution of CHF parameters on a per account basis. Lastly, the miner verifies the modified \(TxToUpdate\) yield the same hash as before the modification. This is due to the hash collision the client created before and ensures hash consistency. If all these checks pass, the \(UpdateTx\) is processed.

\subsection{Processing UpdateTx} \label{processing_update_tx}
After a successful validation of an \(UpdateTx\), the actual update process is performed by miners. In Bazo, only the TX hash is stored in a block, whereas the TX itself is stored in the miner's local storage. This is due to BC-size optimizations and has yet another beneficial side-effect. As the TX hash is not affected by the modification (because of the hash collision), it does not affect the block's Merkle root where the TX originally was included. Thus, it does not affect the hash consistency of BC. This means that miners do not need to modify the original block. The only modification they need to perform is to update the TX in their local storage. To that end, the miner queries the \(TxToUpdate\) from the local storage and performs the modifications on it. First, the new check string is copied over from the \(UpdateTx\) to the \(TxToUpdate\). Subsequently, the new data is copied over from the \(TxToUpdateData\) field of the \(UpdateTx\) to the \(TxToUpdate\)'s data field. Finally, the miner stores and overwrites the modified TX in the local storage.

\subsection{Adding UpdateTx to A New Block} \label{adding_update_tx_to_new_block}
After an \(UpdateTx\) is validated and processed, the miner stores the \(UpdateTx\) in its local storage and adds it - like any other TX - to a new block. As a consequence, the local storage and block data structure of Bazo BC extended in this work. In Bazo, on one hand, a miner's local storage is based on a file database, segmented into buckets where each bucket provides a key/value storage. To store and query \(UpdateTx\), a new bucket is added with the \(UpdateTx\) hash as key and the TX as value. On the other hand, the block data is extended with the following headers to store information about the new \(UpdateTx\).

\begin{itemize}
    \item \textbf{NrUpdateTx} Checksum counter storing the number of \(UpdateTx\) contained in this block.
    \item \textbf{UpdateTxList} Array storing the hashes of the \(UpdateTx\) contained in this block. The hash can be used to query the update TX from the local storage.
\end{itemize}

This work's modified data structure allows each update request to be recorded in a transparent and traceable way. From the content of the \(UpdateTx\) one can derive \textit{(i)} the user who requested a modification, \textit{(ii)} the content of the modification, and \textit{(iii)} the TX on which the modification was applied.

\subsection{Distributing UpdateTx} \label{distributing_update_tx}

Miners in Bazo use a shared Peer-to-peer (P2P) interface to fetch and distribute TXs and blocks among each other. In order to broadcast an \(UpdateTx\) across the miners, the P2P component of the Bazo miner application is extended in this work. At first, new request types were defined as follows. 
\begin{itemize}
    \item \textbf{ReqUpdateTx} Code indicating the P2P request is about fetching an \(UpdateTx\).
    \item \textbf{ResUpdateTx} Code indicating the P2P response is about an \(UpdateTx\) response.
\end{itemize}
Secondly, a new mechanism is implemented for miners to asynchronously fetch \(UpdateTx\) from their peers.

Thus, a new miner can join the BC network by syncing with other ones and building its own local copy of the BC. To that end, the joining miner requests all previous TXs and blocks from its peer miners and validates them in sequential order. This miner will receive all the previous TXs from the network. Moreover, this miner will receive the already updated TXs, and the requests caused the previous updates, too. Thus the joining miner is updated, and no modifications need to be done on this miner's side.

\subsection{TX Aggregation Compatibility} \label{tx_aggregation_compaitibility}
Requirement (7) (\cf Section \ref{discussion}) states that existing BC optimizations shall not be compromised by the system. In the case of Bazo, TX aggregation on funds TXs is implemented to decrease the overall BC size \cite{Bazo-ACM}. In TX aggregation, multiple TXs are summarized and aggregated into one TX by either the sender or receiver, grouped and contained in a designated \(AggTx\). For instance, BC client A sends 5 coins to client B and 7 coins to client C. The aggregation processes combines these TXs as follows. 
\begin{equation}
    FundsTx: A \xrightarrow{} B : 5
\end{equation}
\begin{equation}
    FundsTx: A \xrightarrow{} C : 7
\end{equation}
will be aggregated to
\begin{equation}
    AggTx: A \xrightarrow{} [B,C] : 12
\end{equation}

\(AggTx\) contains the hashes of \(FundsTx\) and \(FundsTx\) and will be added to a new block. To use less space on the BC, \(FundsTx\)  and \(FundsTx\) are removed from the blocks they were included in. Removing \(FundsTx\) and \(FundsTx\) has the consequence that the Merkle root of the affected blocks will change and thus hash consistency will be destroyed. To mitigate this side effect, blocks are connected by an additional fallback link. The fallback link doesn't include the block's Merkle root and thus is not affected by the TX removal \cite{Bazo-ACM}.
In this case, an \(UpdateTx\) enables authorized clients to alter their \(FundsTx\) data field without changing the hash. In Bazo, \(AggTx\) only contains the hash of the aggregated fund TXs, thus, a modification on a TX level will have no effect on the \(AggTx\). 

However, the aggregation of \textit{DataTX}s has to be done differently. In this case, the aggregation shall happen based on the data in the data field and sender address. Otherwise, combination of DataTXs with different data causes data loss.  Which means data1 and data2 sent to B and C as follows, could be aggregated only if data1 = data2. Hence the aggregated data \ie data3 has to be equal to data in all of the aggregated DataTXs.
\begin{equation}
    DataTx: A \xrightarrow{} B : data1
\end{equation}
\begin{equation}
    DataTx: A \xrightarrow{} C : data2
\end{equation}
will be aggregated to
\begin{equation}
    AggTx: A \xrightarrow{} [B,C] : data3
\end{equation}

\subsection{Blockchain Explorer}
\begin{figure}[h]
    \centering
    \includegraphics[width=0.48\textwidth, height=0.18\textheight]{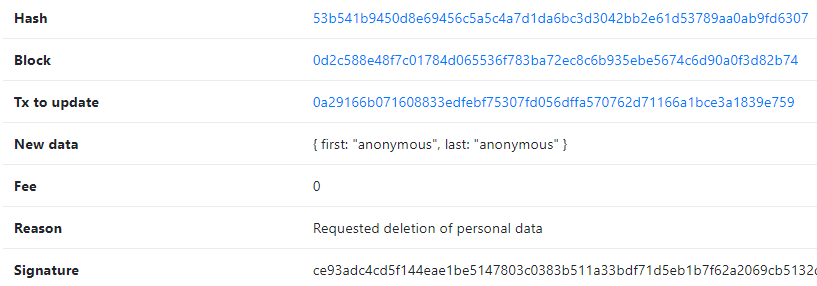}
    \caption{Representation of Data Update Transaction in Blockchain Explorer}
    \label{fig:updatetx}
\end{figure}

For the convenience of users, a BC explorer is adapted in this work \cite{blockExplorer} to illustrate the update TXs as shown in Figure \ref{fig:updatetx}. In the information shown for an example Update TX, the ``new data" field contains the updated data (the identity of a user which is updated to ``anonymous"). Moreover, the ``Reason" of change is now added by users along with the other fields. In this example case, the fee of update TX is set to Zero, due to running in a test BC network. In a public setting, this amount has to be set to a larger than Zero amount to avoid update spams.

\section{Discussion of Evaluations}
\label{sec:Eval}
Fulfilment of the identified requirements in Section \ref{discussion} with the designed and developed approach in this work are elaborated as follows.

\textbf {Requirement (1):} The implementation in this work satisfies the requirement for update and erasure and BC users are able to request the update or erasure of their data by creating an \(UpdateTx\). However, the fact that \(UpdateTx\) carries the \(TxToUpdateData\) field, as well as the fact that updates are mined into the BC yields a corner case where GDPR compliance is hard to assess. 

Assume a user (Alice) issues an \(AccountTx\) with \verb|{name: "Alise"}| included in the \(Data\) field. Unfortunately, the user made a typo and subsequently issues an \(UpdateTx\) to correct the mistake and specifies \(TxToUpdateData\) as \verb|{name: "Alice"}|. After the update is performed, the BC contains the corrected \(AccountTx\), as well as the \(UpdateTx\). If Alice decides to leave the BC later and requests the deletion of her personal data she needs to create another \(UpdateTx\) to erase the \(Data\) field on the \(AccountTx\). While her personal data is removed from the\(AccountTx\), the \(TxToUpdateData\) field of the first \(UpdateTx\) still contains \verb|{name: "Alice"}|. This yields a personal data record (in this case the name used by the user) to remain in the BC, which can't be erased as it is not stored in an editable \(Data\) field but in a TX that must be stored for the tracking of changes purposes. Thus, the requirement (1) is partially satisfied but this special case --which relates more to user action tracks in BC than his/her data-- needs to be addressed in a future work.

\textbf {Requirement (2):} The distribution of the CHF parameters on an account basis, as well as the various validation steps by miners ensure that users only update their own TXs. If a user `\textit{A}' tries to update a TX of user `\textit{B}', the CHF parameters of user `A' would not yield a hash collision for the TX. This is due to the fact that TX was not created by user \textit{A}'s CHF parameters. Thus, requirement (2) is addressed.
 
\textbf {Requirement (3):} As updates are executed in the form of TXs and are mined in blocks, update performance depends on various configurations of the BC such as their scalability and consensus mechanism efficiency. For instance, difficulty-level of the Proof-of-Work puzzle, and block interval impact the overall TX throughput of a BC and so the updating processes. In the developed approach in this work, updates are performed in less than 1 minute. Thus, requirement (3) is fulfilled. 

\textbf {Requirement (4):} The current implementation is self-contained and does not depend on any external systems or services. Thus, requirement (4) is addressed.

\textbf {Requirement (5):} In this work, modifications are performed strictly on a TX-level. Thus, the requirement (5) is addressed.

\textbf {Requirement (6):} After an update is executed, the \(UpdateTX\) is stored in the BC. This design choice yields that \(UpdateTX\) are treated like any other TX in the BC and thus profit from a high level of traceability and auditability. Thus, requirement (6) is addressed.

\textbf {Requirement (7):} The fact that hash collisions enable TX modifications, complement the existing design of Bazo very well. As TXs are stored and updated in the local database of the miners, a modification on the block is obsolete, because blocks only store the TX hash. As a consequence, updates are performed in a non-invasive way, meaning the BC itself is not affected by the update. This enables existing BC optimizations --- such as TX aggregation --- to be unaffected by the updates as well. Thus, requirement (7) is satisfied.

\section{Summary and Conclusions}
\label{sec:summary}

This work set forth the requirements for a successful GDPR compliance of a BC design storing IoT data. 
Accordingly, the designed and implemented approach presented covered the adaptation of a PoS-based and IoT-oriented BC \ie Bazo, at its core cryptographic functions, TXs, algorithms, and processes. The approach developed encompasses BC clients and miners' internal processes in order to become GDPR compliant. 

As a result, a secure and transparent modification of IoT data in a BC is made possible. Via this approach, data modifications are conducted in a fine-grained fashion, \ie at the TX level and not at the block level. By allowing such on-chain modifications of TXs only by TX initiators, \ie data owners, full control and ownership of data is provided to BC clients. The CHF-based processes in this work are distinguished explicitly from related work, since CHF-based processes keep private keys at BC clients' side, thus, offering full authority to BC clients. This approach does not rely on any centralized entity for data storage or modification in contrast to related work analyzed. Furthermore, data update requests are stored in blocks, like all other TXs, such that the traceability of modifications is fully provided.

\section{Acknowledgements}
Firstly, the authors would like to extend many thanks to Bernardo Magri for his collaboration. Secondly, this paper was partially supported by (a) the University of Zürich (UZH), Switzerland and (b) the European Union Horizon 2020 Research and Innovation Program under grant agreement No. 830927, namely the Concordia project. 
\bibliographystyle{IEEEtran}
\balance
\bibliography{main}

\end{document}